\documentclass[a4paper,10pt]{article}

\date{}

\newtheorem{theorem}{Theorem}[section]
\usepackage{braket}

\usepackage{amssymb,latexsym,amsmath,amsfonts}
\usepackage{graphicx}
\usepackage{subfigure}
\usepackage{color}
\usepackage[english]{babel}
\usepackage{authblk}
\usepackage[utf8]{inputenc}
\usepackage{pdfpages}

\usepackage[a4paper,top=2cm,bottom=4cm,left=3.5cm,right=3cm]{geometry} 
\usepackage{fancyhdr}
\usepackage{stmaryrd}
\usepackage{multicol}
\usepackage{lipsum}
\usepackage{enumitem}

\title{Hopf bifurcation analysis of the generalized Lorenz system with time
delayed feedback control}
\author{R. Barresi$^{1}$, M.C. Lombardo$^{1}$, M. Sammartino$^1$       \\
      \tiny{$^2$Dept of Mathematics and Computer Science,  University of Palermo\\ Via Archirafi 34, 90123
 Palermo, Italy.\\
            }
       
       \small{\textit{Email addresses:}\\
       rachele.barresi@unipa.it (R.B.),\\
       mariacarmela.lombardo@unipa.it (M.C.L.),\\
       marcomarialuigi.sammartino@unipa.it (M.S.)}}

\begin{document}

\maketitle

\begin{abstract}
In this work we propose a feedback approach to regulate the chaotic behavior of the whole family of the generalized Lorenz system, by designing a nonlinear delayed feedback control. We first study the effect of the delay on the dynamics of the system and we investigate the existence of Hopf bifurcations. Then, by using the center manifold reduction technique and the normal form theory, we derive the explicit formulas for the direction, stability and period of the periodic solutions bifurcating from the steady state at certain critical values of the delay.

\end{abstract}

\section{Introduction}
Because the fact that several dynamical systems exhibit a chaotic behavior, there has been much interest in the study of chaos. In recent years, the trend of analysing the chaos moved to the new phase consisting of its control and utilization: this means on one hand to design suitable controls to eliminate the chaos, and on the other hand to generate it intentionally.
Our goal in this work is to carry out a rigorous mathematical analysis of dynamic behavior of the whole family of the generalized Lorenz system in its chaotic regime by using time delayed feedback controlling forces in the same spirit of \cite{Gao}. The controller is a nonlinear function of the state variables of the system, therefore the results obtained in this paper can be considered in some way an improvement of the results of \cite{Lorenz_gen}, where the authors study the generalized Lorenz system with a linear version of the control proposed here. The global dynamics of the system depends on the parameter $\alpha\in\left[0,1\right]$, which characterizes the particular system of the whole family. In particular, we obtain the Lorenz system for $\alpha=0$ and the Chen system for $\alpha=1$. Indeed, the systems belonging to the family of the generalized Lorenz system have a similar mathematical structure but they are not topologically equivalent. 
\\In the first part of the paper we introduce the controlled system and we consider the effects of the delay on the steady states also investigating the occurrence of stability switches. Then we show the existence of Hopf bifurcations and so we estabilish that bifurcating periodic solutions exist for each value of the parameter $\alpha \in \left[0,1\right]$. In the last section of the paper we use the same technique pointed out in \cite{Hassard}, based on the center manifold reduction and the normal form theory, in order to determine the direction, stability and period of these periodic solutions which bifurcate from the steady state. This strategy permits to derive the explicit formulas for the properties of the Hopf bifurcation. Moreover, we give numerical simulations of the controlled system, which indicate that when the delay passes through certain critical values, the chaotic behavior is converted to stable periodic orbit for thw whole family of systems.

\section{Nonlinear delayed feedback control for the generalized Lorenz system}
\label{sec:1}
The generalized Lorenz system is described by the following system of ordinary differential equations for the state variables \textit{x ,y, z}:
\begin{equation} \label{eq:gener}
\left\{\begin{array}{l} \frac{dx}{dt} =\left(25\alpha+10\right)\left(y-x\right)\\
\frac{dy}{dt} =\left(28-35\alpha\right)x-xz+\left(29\alpha-1\right)y\\
\frac{dz}{dt} =xy-\frac{\alpha+8}{3}z\end{array}\right.
\end{equation} 
where $\alpha\in \left[0,1\right]$, and it has been introduced for the first time in \cite{LC_gener}. By varying the parameter $\alpha$, we obtain different but structurally similar systems: precisely, they have the same kind of equilibria stability, but they are not topologically equivalent. In particular, the system reduces to the Lorenz system for $\alpha=0$, to the Lu system for $\alpha=0.8$ and to the Chen system for $\alpha=1$, in their chaotic regime.    
The system $\ref{eq:gener}$ has the following three equilibrium points:
\begin{center}
$E_{0}\equiv\left(0,0,0\right)$;    $E_{\pm}\equiv\left(\pm\sqrt{\left(8+\alpha\right)\left(9-2\alpha\right)},\pm\sqrt{\left(8+\alpha\right)\left(9-2\alpha\right)}, 27-6\alpha\right)
$
\end{center} 
and they are all unstable for all $\alpha\in \left[0,1\right]$. In the chaotic regime, the system exhibits an irregular dynamics which makes its evolution unpredictable. For this reason, our aim is to design a suitable control which regulates the system behaviour to any given point of the form \textbf{$x_{p}$}=$\left( x_{r},  x_{r}, 3{x_{r}}^{2}b^{-1}\right)$, that is the form of the two nontrivial fixed points $E_{\pm}$ of the uncontrolled system $\ref{eq:gener}$. Namely, by designing the control 
\begin{center}
 $u=-rx++xz-\gamma y-\sigma\left(y-x_r\right)$, 
\end{center}
where 
\begin{center}
$\sigma=25\alpha+10$, \space $r=28-35\alpha$, \space $b=\frac {\alpha+8}{3}$, \space $\gamma=29\alpha-1$,
\end{center}
the above system is transformed into the closed-loop one:
\begin{equation} \label{eq:c-l system}
\left\{\begin{array}{l} \frac{d\left(x-x_r\right)}{dt} =-\sigma\left(x-x_r\right)+\sigma\left(y-x_r\right)\\
\frac{d\left(y-x_r\right)}{dt} =-\sigma\left(y-x_r\right)\\
\frac{dz}{dt} =xy-\beta z\end{array}\right.
\end{equation}
and it can be proved that $x$ and $y$ both converge to $x_r$ whereas $z$ converges to $\beta^{-1}{x_{r}}^2$. 
\\Unfortunately, the above proposed control does not take into account that the feedback physically enters into the system at a later time, thus in order to avoid this drawback, we consider the delayed feedback controller
\begin{center}
$u=-rx\left(t-\tau\right)+x\left(t-\tau\right)z\left(t-\tau\right)-\gamma y\left(t-\tau\right)-\sigma\left[y\left(t-\tau\right)-x_{r}\right]$
\end{center}
where $\tau$ is the time lag. Since the delay may be destabilizing, our aim is to investigate the stability of the resulting delayed system:
\begin{equation} \label{eq:gener_control}
\left\{\begin{array}{l} \frac{dx}{dt} =\sigma\left(y-x\right)\\
\frac{dy}{dt} =r\left[x-x\left(t-\tau\right)\right]-\left[xz-x\left(t-\tau\right)z\left(t-\tau\right)\right]+\gamma \left[y-y\left(t-\tau\right)\right]-\sigma\left[y\left(t-\tau\right)-x_{r}\right]\\
\frac{dz}{dt} =xy-bz\end{array}\right.
\end{equation}
The characteristic equation associated with the linearization of system $\ref{eq:gener_control}$ around the point \textbf{$E_{+}$} is:
\begin{equation}
 \label{eq:charact_lambda}
\begin{split}
W\left(\lambda\right)\equiv & P\left(\lambda\right)+Q\left(\lambda\right)e^{-\lambda \tau}=\left[\lambda^{3}+\lambda^{2}\left(b+\sigma-\gamma\right)+\lambda\left(\sigma b+K_{1}\right)+\sigma K_{2}\right]+\\
& +\left[\lambda^{2}\left(\sigma+\gamma\right)+\lambda\left(\sigma b+\sigma^{2}-K_{1}\right)+\left(\sigma^{2}b-\sigma K_{2}\right)\right]e^{-\lambda \tau}=0
\end{split}
\end{equation}
where $K_{1}=x_{r}^{2}+\sigma b^{-1}x_{r}^{2}-\sigma r-\gamma\left(\sigma+b\right)$ and $K_{2}=3x_{r}^{2}-b\gamma-br$.
\\Without any delay $\left(\tau=0\right)$, Eq. $\ref{eq:charact_lambda}$ becomes 
\begin{equation}\label{eq:char_tauzero}
\lambda^{3}+\lambda^{2}\left(b+2\sigma\right)+\lambda\left(2\sigma b+\sigma^{2}\right)+\sigma^{2}b=0
\end{equation}
and, noticing that 
\begin{center}
$
b+2\sigma>0, \qquad \sigma^{2}+2b\sigma>0,\qquad \sigma^{2}b>0,$\\$
\left(b+2\sigma\right)\left(\sigma^{2}b+2b\sigma\right)-\sigma^{2}b=2\sigma\left(\sigma +b\right)^{2}>0,
$
\end{center}
and by the Routh-Hurwitz criterion, Eq. $\ref{eq:char_tauzero}$ has three roots with negative real part for all $\alpha\in\left[0,1\right]$, as aspected since the control stabilizes the system \cite{Benard}. Thus we have the following result.




\begin{theorem}
The equilibrium point $E_{+}$ of system $\ref{eq:gener}$ is globally asymptotically stable when $\tau=0$, for all $\alpha\in\left[0,1\right]$.
\end{theorem} 
\textbf{Proof.} The equilibrium $E_{+}$ is locally asymptotically stable when $\tau=0$ because all the roots of the characteristic equation have negative real part. To prove the global stability, we  consider a Lyapunov functional $L:\mathbb{R}^{2}\rightarrow\mathbb{R}$ given by:
\begin{center}
$
V\left(\tilde{x},\tilde{y}\right)=2\tilde{x}^2+3\tilde{y}^2+2\tilde{x}\tilde{y}=\left(\tilde{x}+\tilde{y}\right)^2+2\tilde{y}^2+\tilde{x}^2 ,
$
\end{center}
where $\tilde{x}=x-x_r$ and $\tilde{y}=y-x_r$, and this functional is positive definite $\forall\left(\tilde{x},\tilde{y}\right)\neq\left(0,0\right)$ and $V\left(0,0\right)=0$. In the same way, it's easy to show that $\dot{V}$ is negative definite $\forall\left(\tilde{x},\tilde{y}\right)\neq\left(0,0\right)$. By the Lyapunov stability theorem, we can conclude that the equilibrium point $E_{+}$ is globally asymptotically stable.
\hspace{\stretch{1}} $\Box$ 
\vspace{0.5cm}
\\Time delays are known to cause destabilization of equilibria and produce oscillations through Hopf bifurcations. Moreover, it has been observed that further increase in the delay may result in restabilization. This phenomenon is called stability switch: by increasing the delay $\tau$, it can occur that zeroes of the characteristic equation $\ref{eq:charact_lambda}$ cross the imaginary axis and the system may change from stable to unstable or vice versa. In order to discuss the existence of such phenomena in system $\ref{eq:gener_control}$, we look at the characteristic equation $\ref{eq:charact_lambda}$ as a function of the delay $\tau$ and examine the location of roots and the direction of motion as they cross the imaginary axis \cite{Zeros_trasc}. 
\\It's crucial to determine the condition to obtain destabilization, that is the critical value $\tau_c$ at which there is the existence of purely imaginary characteristic values. Indeed, if the roots of $\ref{eq:charact_lambda}$ are in the left-half plane $\forall\tau\geq 0$, the equilibrium is asymptotically stable for all $\tau$. Otherwise, there could be values of $\tau$ for which a pair of complex conjugate roots of $\ref{eq:charact_lambda}$ crosses the imaginary axis, and if the cross is from left to right, the equilibrium is destabilized, instead if the cross is from right to left an unstable equilibrium is stabilized when $\tau$ increases. 
\\Following the ideas of \cite{Zeros_trasc,Discrete_delay} and making use of Theorem 1 \cite{Zeros_trasc}, we assume that $\lambda=i\nu$, $\nu>0$, is a root of $\ref{eq:charact_lambda}$ for some positive $\tau$ and we define the auxiliary function 
\begin{equation}\label{eq:F_x}
\begin{split}
&F\left(x\right)=\left\|P\left(i\nu\right)\right\|^{2}-\left\|Q\left(i\nu\right)\right\|^{2}=
x^3+x^2\left[b^2-2b\gamma-4\sigma\gamma-2K_1\right]+ \\
&+x\left[2\sigma b\left(2K_1-K_2\right)+\sigma^2\left(2b\gamma+2K_1-4K_2-\sigma^2\right)\right]+b\sigma^3\left(2K_2-b\sigma\right).
\end{split}
\end{equation}
where $P$ and $Q$ have been already defined in $\ref{eq:charact_lambda}$ and $x=\nu^2$.
\\It is clear that the existence of purely imaginary eigenvalues for system $\ref{eq:gener_control}$ is equivalent to the existence of positive roots of $F$. If $F$ has a positive simple root $x_0$, then there exists a pair of $\pm i\nu_0$ of purely imaginary eigenvalues with $\nu_0=\sqrt{x_0}$ and, for this $\nu_0$, we find a sequence of $\left\{{\tau_0}^n\right\}$ of delays for which stability switches can occur (at least a finite number); furthermore, there exists a positive $\tau_c$ such that the system is unstable for all $\tau>\tau_c$.
\\Note that
\begin{center}
$
F'\left(x\right)=3x^2+2x\left[b^2-2b\gamma-4\sigma\gamma-2K_1\right]+\left[2\sigma b\left(2K_1-K_2\right)+\sigma^2\left(2b\gamma+2K_1-4K_2-\sigma^2\right)\right]
$
\end{center}
and
\begin{equation}
\Delta=\left[b^2-2b\gamma-4\sigma\gamma-2K_1\right]^{2}-3\left[2\sigma b\left(2K_1-K_2\right)+\sigma^2\left(2b\gamma+2K_1-4K_2-\sigma^2\right)\right].
\end{equation}
\begin{enumerate}[label=(\alph*)]
\item If $\Delta\leq 0$, then $F'\left(x\right)\geq 0$ and $F\left(x\right)$ is monotonically increasing. Therefore, when $F\left(0\right)\geq 0$ and $\Delta\leq 0$, $F\left(x\right)=0$ has no positive roots and all the characteristic roots will remain to the left of the imaginary axis for all $\tau>0$.
\item If $F\left(0\right)<0$, since $\lim_{x \to \infty}F\left(x\right)=\infty$, there is at least one positive root of $F\left(x\right)=0$ and the characteristic roots can cross the imaginary axis. 
\item If $\Delta>0$, then the graph of $F\left(x\right)$ has critical points 
\begin{center}
$x^{*}=\frac{-\left(b^2-2b\gamma-4\sigma\gamma-2K_1\right)+\sqrt{\Delta}}{3}\qquad x^{**}=\frac{-\left(b^2-2b\gamma-4\sigma\gamma-2K_1\right)-\sqrt{\Delta}}{3}$
\end{center}
and, moreover, if $x^{*}>0$ and $F\left(x^{*}\right)<0$, then $F\left(x\right)=0$ has positive roots \cite{Li_Shu,Song}.
\end{enumerate}


According to Theorem 1 \cite{Zeros_trasc}, stability switches are possible for each positive root $x_j$ of $\ref{eq:F_x}$ and the cross is from left to right  if $F'\left(\nu_0\right)>0$, and from right to left is $F'\left(\nu_0\right)<0$.
\\ The characteristic quasi-polynomial $\ref{eq:charact_lambda}$ for $\lambda=i\nu$ has the form
\begin{equation} \label{eq:W_imaginary}
W\left(i\nu\right)=\alpha_{3}-\alpha_{1}\cos\left(\nu\tau\right)-\alpha_{2} \sin\left(\nu\tau\right)+i\left[\alpha_{4}-\alpha_{2} \cos\left(\nu\tau\right)+\alpha_{1} \sin\left(\nu\tau\right)\right]=0
\end{equation}
where:
\begin{center}
$
\alpha_1= \nu^{2}\left(\sigma+\gamma\right)-\sigma^{2}b+\sigma K_2,\qquad\alpha_2 = -\nu\left(\sigma b+\sigma^{2}-K_1\right),$  \\
$\alpha_3 = -\nu^{2}\left(b+\sigma-\gamma\right)+\sigma K_2,\qquad\alpha_4= -\nu^{3}+\nu\left(\sigma b+K_1\right).$
\end{center}
Let $x_{j}$, $1\leq j\leq 3$, be a positive root of $F\left(x\right)=0$, and $\nu_{j}=\sqrt{x_{j}}$. Then $\nu_j$ satisfies $\ref{eq:W_imaginary}$, that is equivalent to the following system:
\begin{equation} \label{eq:system_alpha}
\left\{\begin{array}{l} \alpha_{3}-\alpha_{1}\cos\left(\nu\tau\right)-\alpha_{2}\sin\left(\nu\tau\right)=0\\
\alpha_{4}-\alpha_{2}\cos\left(\nu\tau\right)+\alpha_{1}\sin\left(\nu\tau\right)=0\end{array}\right.
\end{equation} 
By setting
\begin{equation} 
P\left(\i\nu\right)=P_R\left(i\nu\right)+iP_I\left(i\nu\right), \qquad Q\left(\i\nu\right)=Q_R\left(i\nu\right)+iQ_I\left(i\nu\right)
\end{equation} 
where
\begin{center}
$
P_R\left(i\nu\right)=-\nu^2\left(\sigma+b-\gamma\right)+\sigma K_2 \qquad P_I\left(i\nu\right)=-\nu^3+\nu\left(\sigma b+K_1\right)
$\\$
Q_R\left(i\nu\right)=-\nu^2\left(\sigma+\gamma\right)+\sigma^2 b-\sigma K_2 \qquad Q_I\left(i\nu\right)=\nu\left(\sigma b+\sigma^2-K_1\right)
$
\end{center}
after simplification the above system implies that
\begin{equation} \label{eq:sincos_alpha}
\sin\left(\nu\tau\right)=\frac{-P_R Q_I+Q_R P_I}{Q^{^2}_R+Q^{^2}_I}, \qquad \cos\left(\nu\tau\right)=-\frac{P_R Q_R+P_I Q_I}{Q^{^2}_R+Q^{^2}_I}
\end{equation} 
Thus, for each positive root $\nu_j$, it yields the following sequence of delays $\left\{\tau^{n}_j\right\}$ for which there are pure imaginary roots of $\ref{eq:charact_lambda}$:
\begin{equation}\label{eq:tau_n}
\tau^{n}_j=\frac{1}{\nu_j}\left\{\arctan\left(\frac{-P_R Q_I+P_I Q_R}{-\left(P_R Q_R+P_I Q_I\right)}\right)+2n\pi\right\}, \qquad \mathrm{for\hspace{1mm}n=0,1,} \dots
\end{equation}

\begin{figure*}[h]
\begin{center}
\includegraphics[width=8cm,height=6cm]{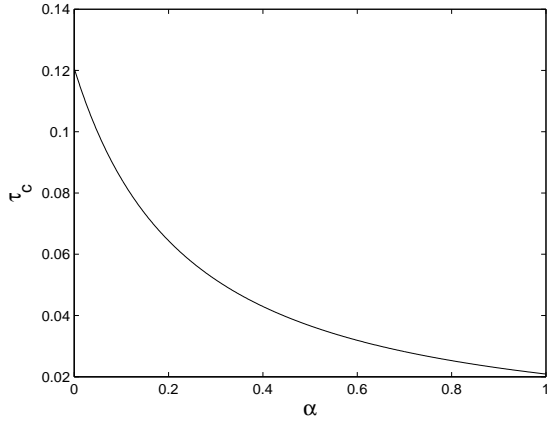}
\caption{Plot of $\tau_c$ as a function of $\alpha$.}
\label{fig:tau_c_alpha}
\end{center}
\end{figure*}

We numerically find that for all $\alpha\in\left[0,1\right]$ there exist three real roots of $\ref{eq:F_x}$, of which only two are positive, $\nu_-<\nu_+$, and  crossing is from left to right with increasing $\tau$ occurs whenever $\tau$ assumes a value corresponding to $\nu_+$, as $F'\left(\nu_+\right)>0$, and $F'\left(\nu_-\right)<0$ so crossing from right to left occurs for values of $\tau$ corresponding to $\nu_-$. Moreover, since the zero solution is stable for $\tau=0$, then $\tau^{0}_+<\tau^{0}_-$. We observe that
\begin{center}
$\tau^{j+1}_+-\tau^{j}_+=\frac{2\pi}{\nu_+}<\frac{2\pi}{\nu_-}=\tau^{j+1}_--\tau^{j}_-$
\end{center}
therefore there can be only a finite number of stability switches, if they occur. In our case, we see that for all $\alpha\in\left[0,1\right]$ the smallest value $\tau^{0}_+$ is the critical value $\tau_c=\tau^{0}_+$ at which stability switch occurs from stable to unstable, so that the stability is lost at $\tau=\tau_c$ and for $\tau>\tau_c$ the solution remains unstable.  
\\ \textit{Remark}. By solving system $\ref{eq:system_alpha}$ numerically, we find for each value of $\alpha$ the critical delay $\tau_c$, in particular $\tau_c\approx 0.122$ for $\alpha=0$, $\tau_c\approx 0.0253$ for $\alpha=0.8$ and $\tau_c\approx 0.021$ for $\alpha=1.0$. Actually, as we can see in Fig. $\ref{fig:tau_c_alpha}$, $\tau_c$ is a decreasing function of $\alpha$ on the interval $\left[0,1\right]$. 

\begin{figure*}[h!]
\includegraphics[width=\textwidth,height=6cm]{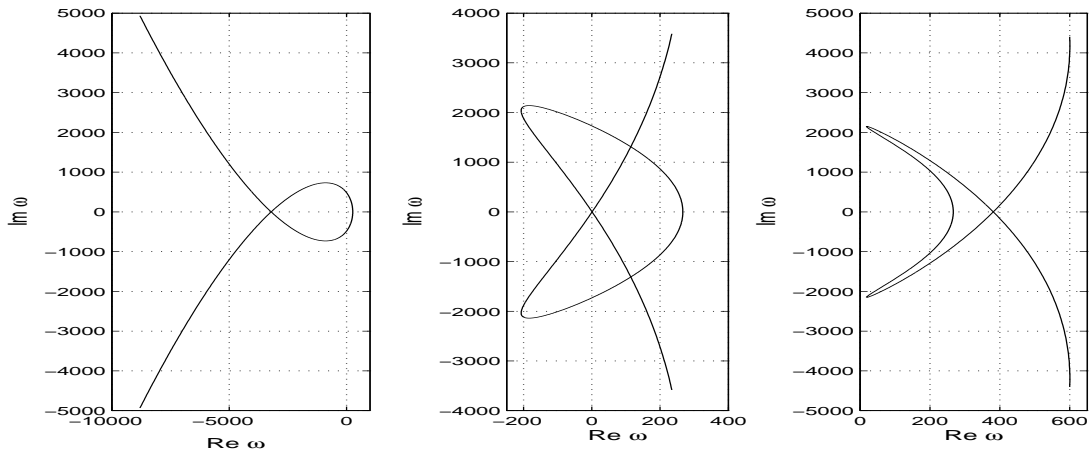}
\caption{Map in the $\omega$-plane for (a) $\tau<\tau_c$, (b) $\tau=\tau_c=0.122$ and (c) $\tau>\tau_c$ in the case $\alpha=0$.}
\label{fig:lorenz_hod}
\end{figure*}

\begin{figure*}[h!]
\includegraphics[width=\textwidth,height=6cm]{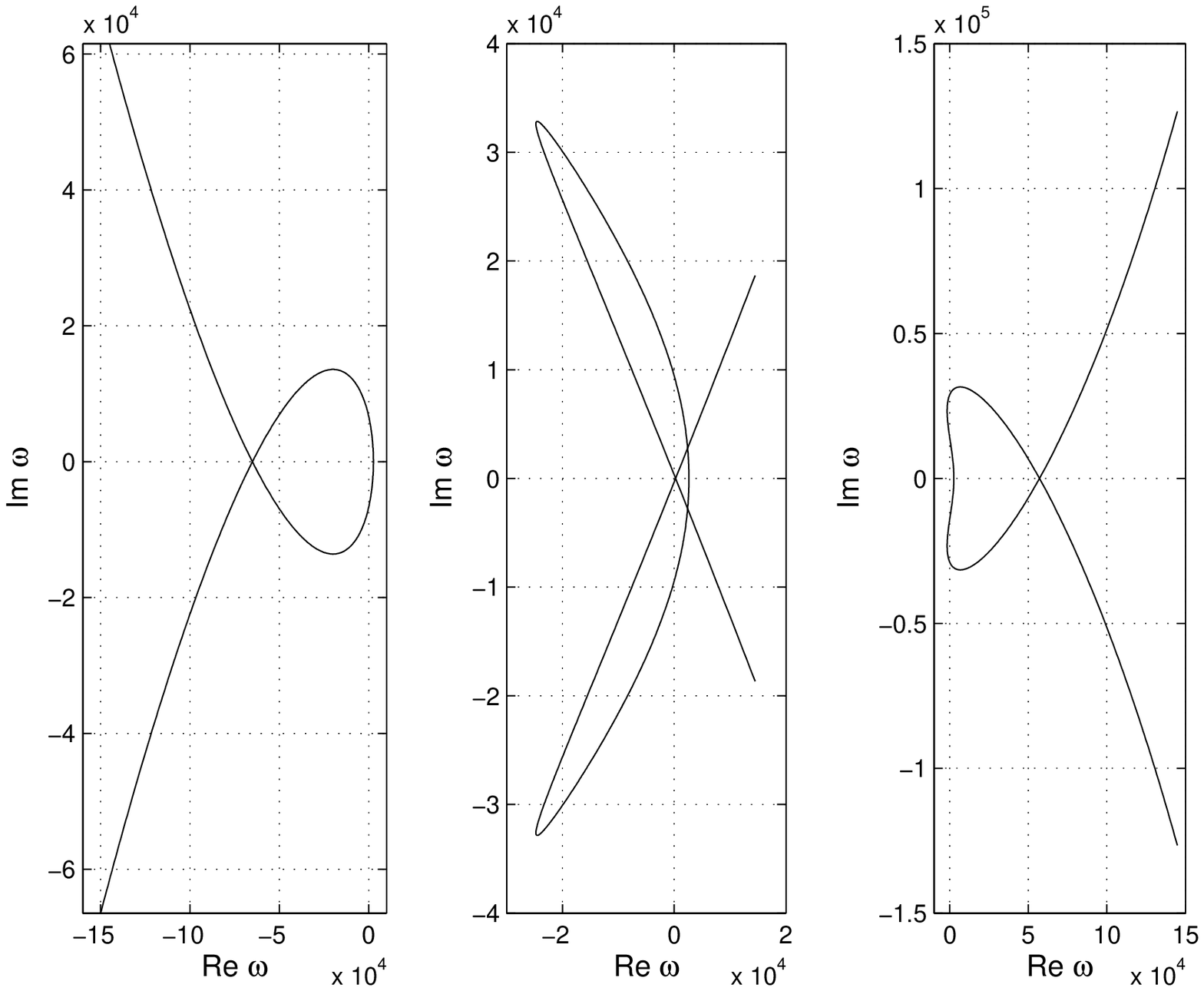}
\caption{Map in the $\omega$-plane for (a) $\tau<\tau_c$, (b) $\tau=\tau_c=0.0253$ and (c) $\tau>\tau_c$ in the case $\alpha=0.8$.}
\label{fig:lu_hod}
\end{figure*}

\begin{figure*}[h!]
\includegraphics[width=\textwidth,height=6cm]{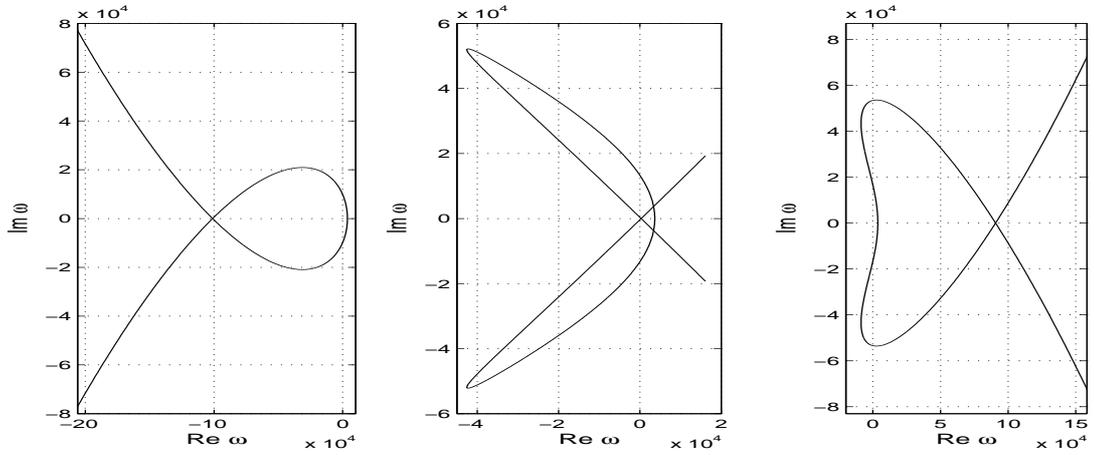}
\caption{Map in the $\omega$-plane for (a) $\tau<\tau_c$, (b) $\tau=\tau_c=0.021$ and (c) $\tau>\tau_c$ in the case $\alpha=1$.}
\label{fig:chen_hod}
\end{figure*}

Therefore, if we consider the trascendental equation $\ref{eq:charact_lambda}$ as a complex variable mapping problem from the $\lambda$-plane to the $\omega$-plane \cite{Murray}
\begin{equation}\label{eq:map_omega}
\omega=W\left(\lambda\right)
\end{equation}
then the critical value $\tau_{c}$ is such that $\omega=0$ has solutions with $Re\lambda>0$ for $\tau>\tau_{c}$. Hence, $\tau_{c}$ is the bifurcation value for which $Re\lambda=0$. Without any delay, we already showed that all the solutions of $\omega=0$ have $Re\lambda<0$. This implies that, if one considers the contour in the $\lambda$-plane consisting of the imaginary axis and of a semi-circle of infinite radius, its image under the map given by $\ref{eq:map_omega}$ does not enclose the origin of the $\omega$-plane (see Figs.$\ref{fig:lorenz_hod}$-$\ref{fig:chen_hod}$(a)). Now, when $\tau>0$, if we consider the mapping $\omega=W\left(\lambda=i\nu\right)$, as soon as the trasformed curve passes through the origin in the $\omega$-plane this gives the critical value $\tau_c$ (see Figs.$\ref{fig:lorenz_hod}$-$\ref{fig:chen_hod}$(b)). For $\tau>\tau_c$, the mapping is shown in Figs. $\ref{fig:lorenz_hod}$-$\ref{fig:chen_hod}$(c) and the origin in the $\omega$-plane is crossed.

\begin{figure*}[h]
\begin{center}
\includegraphics[width=\textwidth,height=12cm]{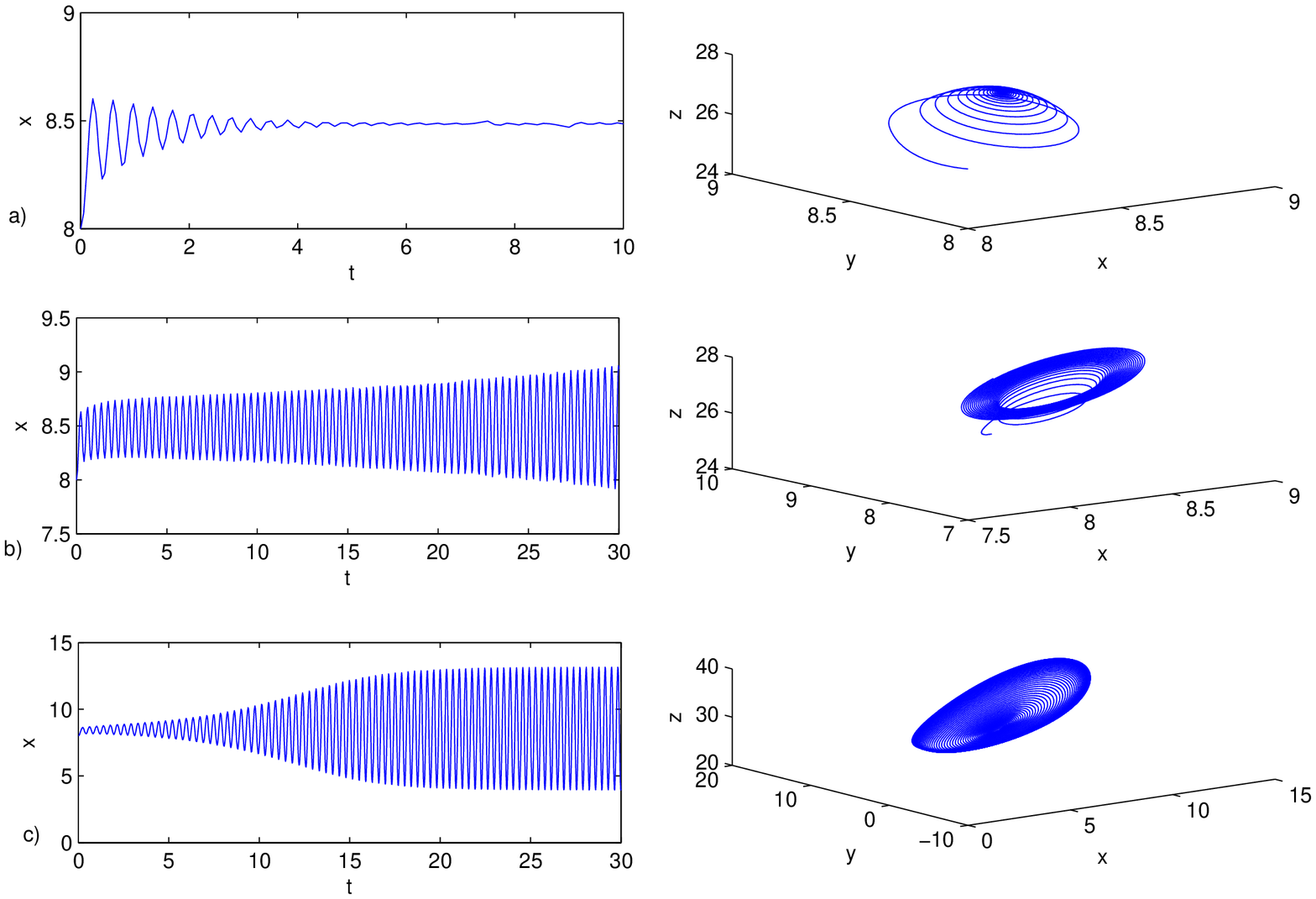}
\end{center}
\caption{Case $\alpha=0$. Solution x(t) of system $\ref{eq:gener_control}$: (a) The system is regulated to the equilibrium point $E_{+}$ at $\tau=0.112$.(b) Destabilization of the steady state occurs at $\tau=\tau_c=0.122$. (c) The solution exhibits oscillations whose amplitude stabilizes at $\tau=0.125$.}
\label{fig:lorenz_sol_x}
\end{figure*}
\begin{figure*}[h]
\begin{center}
\includegraphics[width=\textwidth,height=12cm]{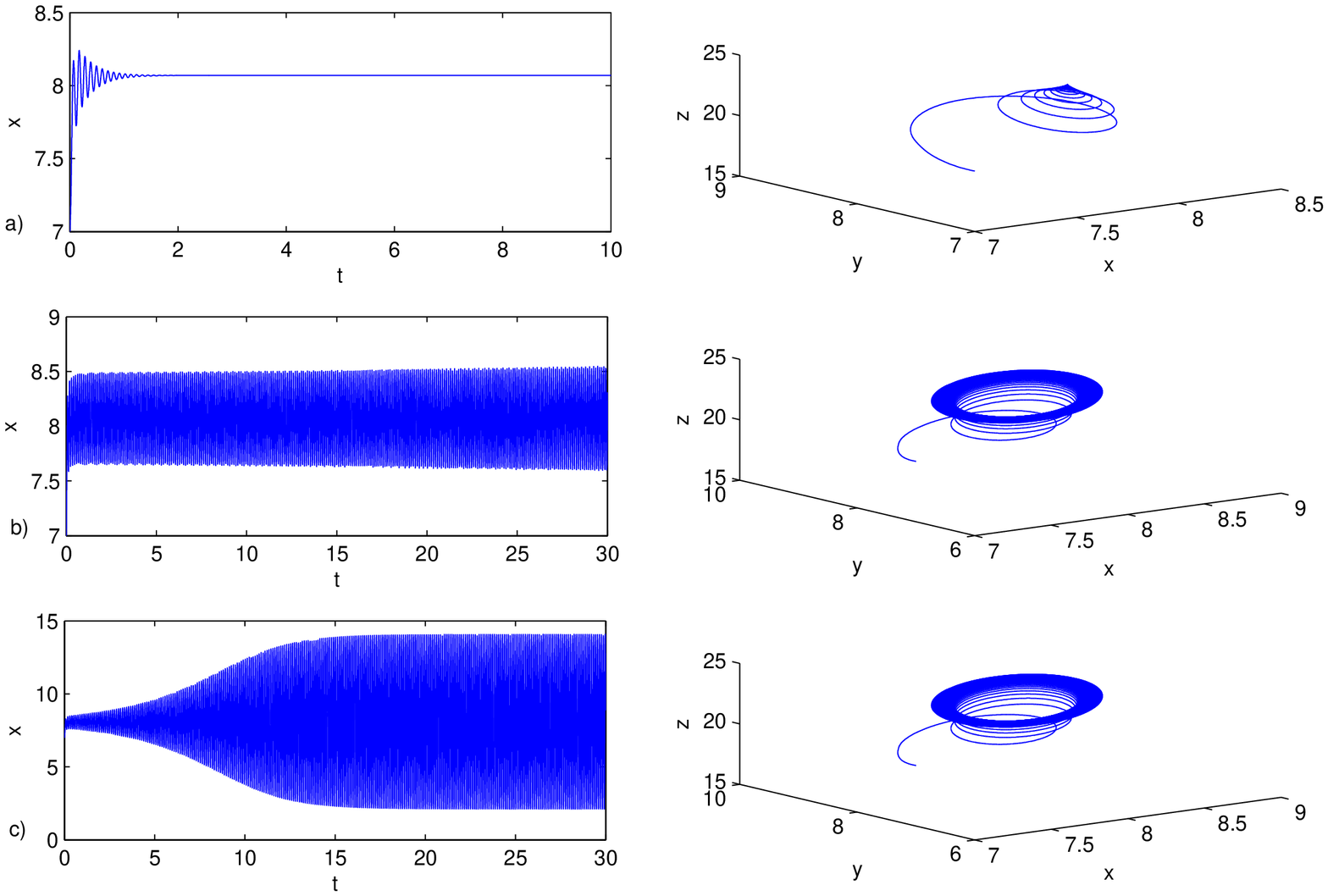}
\end{center}
\caption{Case $\alpha=0.8$. Solution x(t) of system $\ref{eq:gener_control}$: (a)  The system is regulated to the equilibrium point $E_{+}$ at $\tau=0.023$. (b) Destabilization of the steady state occurs at $\tau=\tau_c=0.0253$. (c) The solution exhibits oscillations whose amplitude stabilizes at $\tau=0.0255$.}
\label{fig:lu_sol_x}
\end{figure*}
\begin{figure*}[h]
\begin{center}
\includegraphics[width=\textwidth,height=12cm]{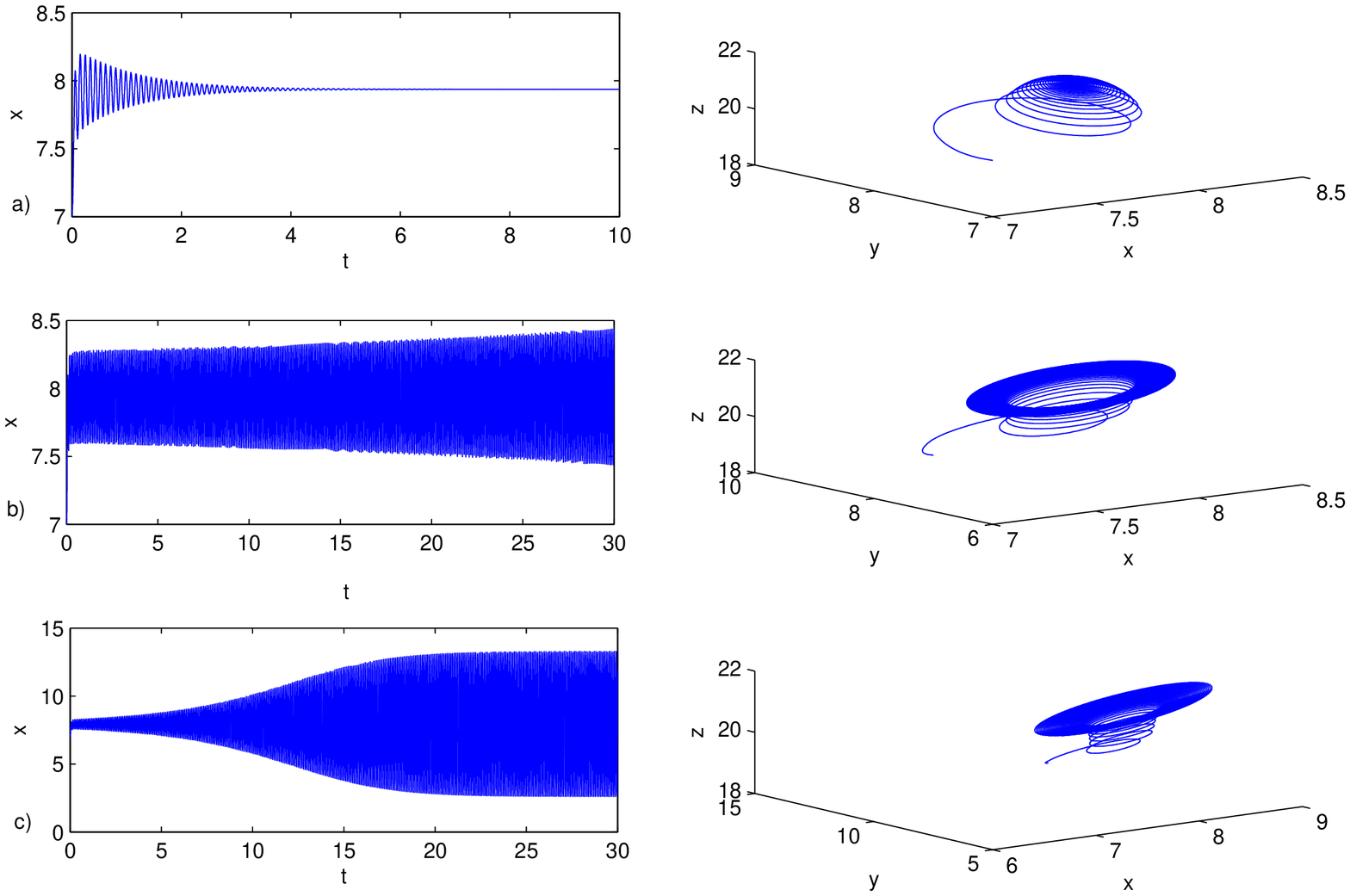}
\end{center}
\caption{Case $\alpha=1$. Solution x(t) of system $\ref{eq:gener_control}$: (a) The system is regulated to the equilibrium point $E_{+}$ at $\tau=0.0204$.(b) Destabilization of the steady state occurs at  $\tau=\tau_c=0.021$. (c) The solution exhibits oscillations whose amplitude stabilizes at $\tau=0.022$.}
\label{fig:chen_sol_x}
\end{figure*}

Numerical simulations of the ODEs system show that for $\tau<\tau_c$ the solution, after some transient oscillations, stabilizes to the equilibrium position (see Figs.$\ref{fig:lorenz_sol_x}$-$\ref{fig:chen_sol_x}$ (a)).
\\However, with an increasing delay, solutions exhibit an oscillatory behavior for all $\alpha\in\left[0,1\right]$, as shown in Figs.$\ref{fig:lorenz_sol_x}$-$\ref{fig:chen_sol_x}$ (b),(c), and this dynamic suggests that the system exhibits Hopf bifurcation, even if there is a qualitative difference due to the fact that the threshold value of delay changes.

We obtain the following result.
\begin{theorem}\label{transv_cond}

\begin{enumerate}[label=(\roman*)]
\item If $\Delta\leq 0$ and $b\sigma^3\left(2K_2-b\sigma\right)\geq 0$, then the equilibrium point $E_+$ remains asymptotically stable for all $\tau\geq 0$.
\item If either  \begin{enumerate}[label=(\alph*)]
\item $b\sigma^3\left(2K_2-b\sigma\right)<0$, or
\item $\Delta>0$, $x^{*}>0$ and $F(x^{*})<0$,
\end{enumerate}
then there exist $\tau_c>0$ and $\nu_0$ as defined above such that the equilibrium point $E_+$ is asymptotically stable for $\tau\in[0,\tau_c)$. Furthermore, if $F'(\nu^{2}_0)\neq 0$, then the system $\ref{eq:gener_control}$ undergoes a Hopf bifurcation at the equilibrium  $E_{+}$ when $\tau=\tau_c$.
\end{enumerate}
\end{theorem} 

\begin{figure*}[h]
\begin{center}
\includegraphics[width=12cm,height=6cm]{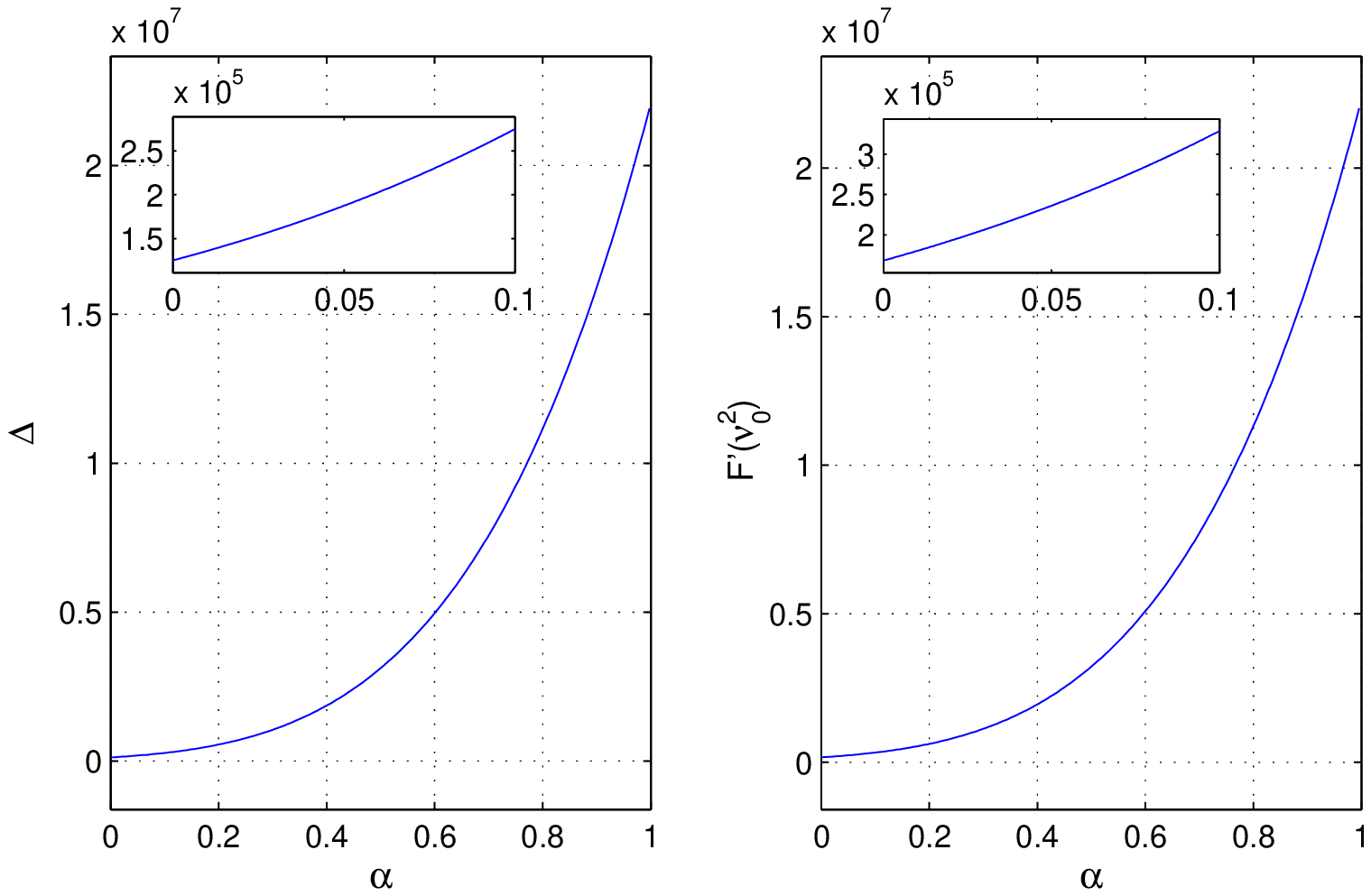}
\end{center}
\caption{Plots of (a) $\Delta$ and (b) $F'(\nu_0^2)$ as functions of $\alpha$. Both the functions are increasing and non zero on the domain $\left[0,1\right]$, as shown in the zoom on the left side on top of the plots.}
\label{fig:f_deriv_delta_alpha}
\end{figure*}

\textbf{Proof.} It remains to show the transversality condition for the Hopf bifurcation holds at $\tau=\tau_c$. 
\\
We first set
\begin{center}
$
a_0=\sigma K_2, \qquad a_1=b\sigma+K_1, \qquad a_2=b+\sigma-\gamma,$
\\
$
b_0=b\sigma^2-\sigma K_2, \qquad b_1=b\sigma+\sigma^{2}-K_1, \qquad b_2=\sigma+\gamma.
$
\end{center}
So, differentiating Eq.$\ref{eq:charact_lambda}$ with respective to $\tau$, we obtain
\begin{equation}
\left[\frac{d\lambda}{d\tau}\right]^{-1}=-\frac{P'(\lambda)}{\lambda Q(\lambda)}+\frac{Q'(\lambda)}{\lambda Q(\lambda)}-\frac{\tau}{\lambda}=\frac{(3\lambda^2+2a_2\lambda+a_1)e^{\lambda\tau}}{\lambda(b_2\lambda^2+b_1\lambda+b_0)}
+\frac{2b_2\lambda+b_1}{\lambda(b_2\lambda^2+b_1\lambda+b_0)}-\frac{\tau}{\lambda}.
\end{equation}
Using Eq.$\ref{eq:W_imaginary}$, we obtain
\begin{equation}
\begin{split}
\left[\frac{dRe(\lambda)}{d\tau}\right]^{-1}_{\tau=\tau_c} &=Re\left[-\frac{P'(\lambda)}{\lambda Q(\lambda)}\right]_{\tau=\tau_c}+Re\left[\frac{Q'(\lambda)}{\lambda Q(\lambda)}\right]_{\tau=\tau_c}\\
&=\frac{3\nu_0^6+2(a_2^2-b_2^2-2a_1)\nu_0^4+(a_1^2-2a_0 a_1-b_1^2+2b_0b_2)\nu_0^2}{b_1^2\nu_0^4+\nu_0^2(b_0-b_2\nu_0^2)^2}\\
&=\frac{F'(\nu_0^2)}{b_1^2\nu_0^2+(b_0-b_2\nu_0^2)^2}.
\end{split}
\end{equation}
Therefore 
\begin{equation}
sign\left[\frac{dRe(\lambda)}{d\tau}\right]_{\tau=\tau_c}=signF'(\nu_0^2).
\end{equation}
If $F'(\nu_0^2)\neq 0$, the transversality condition holds and a Hopf bifurcation occurs at $\tau=\tau_c$.
\hspace{\stretch{1}} $\Box$ 
\vspace{0.5cm}
\\
\textit{Remark.} If $F'(\nu_0^2)\neq 0$, then $\frac{dRe(\lambda\left(\tau_c\right))}{d\tau}\neq 0$. If $\frac{dRe(\lambda\left(\tau_c\right))}{d\tau}<0$, then characteristic equation has roots with positive real parts for $\tau<\tau_c$ and close to $\tau_c$, but this contradicts the fact that $E_+$ is asimptotically stable for $0\leq\tau<\tau_c$ as in Theorem $\ref{transv_cond}$. Thus, $F'(\nu_0^2)\neq 0$, then $\frac{dRe(\lambda\left(\tau_c\right))}{d\tau}>0$.  
\\The expression of $\Delta$ as a function of $\alpha$ is quite cumbersome, but numerical simulations show that $\Delta>0$ for all $\alpha\in\left[0,1\right]$ (see Fig. $\ref{fig:f_deriv_delta_alpha}$(a)), and in the same way we show that $F'(\nu_0^2)\neq 0$ $\forall\alpha\in\left[0,1\right]$ (see Fig.$\ref{fig:f_deriv_delta_alpha}$(b)). Moreover, it's easy to prove that $x^*>0$ and $F\left(x^*\right)<0$, thus the hypothesis ii $\left(a\right)$ is always verified and the system $\ref{eq:gener_control}$ undergoes a Hopf bifurcation $\forall\alpha\in\left[0,1\right]$ at the equilibrium  $E_{+}$ when $\tau=\tau_c$.

Transversality condition defines the direction of motion of $\lambda$ as $\tau$ varies. It is also the necessary condition for the existence of periodic orbits: by varying $\tau$, the critical characteristic roots cross the imaginary axis with nonzero velocity. 
\\In this section, we have identified the conditions under which delay can destabilize the equilibrium and leads to Hopf bifurcation. We have shown that, when $\tau>\tau_c$, periodic solutions exist. Now we have to investigate the direction, stability and period of these periodic solutions bifurcating from the steady state.


\section{Direction and stability of the Hopf bifurcation}
\label{sec:2}
The aim of this section is to derive the explicit formulas determining in particular the direction, stability and period of the periodic solutions bifurcating from the steady state at the critical value $\tau_c$, using the normal form and the center manifold theory pointed out in \cite{Hassard} and \cite{Song}.
\\Throughout this section, we assume that the system $\ref{eq:gener_control}$ undergoes Hopf bifurcation at the steady state $E_{+}=(x^*,y^*,z^*)$ for $\tau=\tau_c$ and that $\pm i\nu_0$ are the corresponding purely imaginary roots of the characteristic equation at the steady state $E_{+}$.
\\We first traslate the equilibrium $E_+$ to the origin through the change of variables $\bar{x}=x-x^*$, $\bar{y}=y-y^*$, $\bar{y}=y-y^*$, and we drop the bars for semplification of notation, thus system  $\ref{eq:gener_control}$ is transformed into
\begin{equation}\label{eq:syst_scal}
\left\{\begin{array}{l} \frac{dx}{dt} =\sigma\left(y-x\right)\\
\frac{dy}{dt} =r\left[x-x\left(t-\tau\right)\right]-\left[xz-x\left(t-\tau\right)z\left(t-\tau\right)+x_r\left(z-z\left(t-\tau\right)\right)+\frac{{x_r}^2}{b}\left(x-x\left(t-\tau\right)\right)\right]+
\\+\gamma \left[y-y\left(t-\tau\right)\right]-\sigma y\left(t-\tau\right)\\
\frac{dz}{dt} =xy+x_r\left(x+y\right)-bz\end{array}\right.
\end{equation}
Let $\tau=\tau_c+\mu$, and we use $\mu$ as the bifurcation parameter with $\mu=0$ the Hopf bifurcation value. We scale the time t$\rightarrow \left(t/\tau\right)$ in system $\ref{eq:syst_scal}$ and set 
\begin{center}
$B_1=
\begin{pmatrix}
-\sigma & \sigma & 0 \\
r+\frac{{x_r}^2}{b} & \gamma & x_r \\
x_{r} & x_{r} & -b
\end{pmatrix}
$
\hspace{0.3cm}
$B_2=
\begin{pmatrix}
0 & 0 & 0 \\
-(r+\frac{{x_r}^2}{b}) & -(\gamma+\sigma) &- x_r \\
0 & 0 & 0
\end{pmatrix}
$
\end{center}
We define an operator $L_\mu:\mathcal{C}\left(\left[-1,0\right],\mathbb{R}^3\right)\rightarrow\mathbb{R}$ as 
\begin{equation}\label{eq:L_mu}
L_\mu\left(\phi\right)=\left(\tau_c+\mu\right)B_1\phi\left(0\right)+\left(\tau_c+\mu\right)B_2\phi\left(-1\right),
\end{equation}
and
\begin{equation}
f\left(\mu,\phi\right)=\left(\tau_c+\mu\right)\begin{pmatrix}
0 \\
-\phi_1(0)\phi_3(0)+\phi_1(-1)\phi_3(-1) \\
\phi_1(0)\phi_2(0)
\end{pmatrix}
\end{equation}
for $\phi=(\phi_1,\phi_2,\phi_3)^{T}\in\mathcal{C}([-1,0],\mathbb{R}^3)$. So, we can rewrite the system $\ref{eq:syst_scal}$ as an FDE in $\mathcal{C}([-1,0],\mathbb{R}^3)$ as
\begin{equation}\label{eq:vdot}
\dot{v}(t)=L_\mu(v_t)+f(\mu,v_t)
\end{equation}
where $v(t)=(x(t),y(t),z(t))^{T}\in\mathbb{R}^3$.
Now, by the Riesz representation theorem, there exists a function $\eta(\theta,\mu)$ of bounded variation for $\theta\in[-1,0]$, such that the operator $L_\mu$ can be represented in an integral form as follows
\begin{equation}\label{eq:L_integ}
L_\mu(\phi)=\int_{-1}^{0} d\eta(\theta,0)\, \phi(\theta), \qquad  for \quad \phi\in\mathbb{C}([-1,0],\mathbb{R}^3)
\end{equation}
and we can choose 
\begin{equation}\label{eq:eta}
\eta(\theta,\mu)=(\tau_c+\mu)B_1\delta(\theta)-(\tau_c+\mu)B_2\delta(\theta+1)
\end{equation} 
where $\delta$ is the Dirac delta function. 
\\The next step is to define operators A and R so that system $\ref{eq:vdot}$ can be written as an abstract ODE in the Banach space $\mathcal{C}^1([-1,0],\mathbb{R}^3)$. So, we define for $\phi\in\mathcal{C}^1([-1,0],\mathbb{R}^3)$
\\
$A(\mu)\phi(\theta) = \begin{cases} \frac{d\phi(\theta)}{d\theta} & \theta\in[-1,0)\\
 \int_{-1}^{0} d\eta(s,\mu)\, \phi(s)& \theta=0\end{cases}
$
\\and
\\
$R(\mu)\phi(\theta) = \begin{cases} 0 & \theta\in[-1,0)\\
 f(\mu,\theta) & \theta=0\end{cases}
$
\\
Then system $\ref{eq:vdot}$ is equivalent to 
\begin{equation}\label{eq:vdot_second}
\dot{v_t}=A(\mu)v_t+R(\mu)v_t.
\end{equation}
where $v_t(\theta)=v(t+\theta)$ for $\theta\in[-1,0]$.
For $\psi\in\mathcal{C}^1([-1,0],(\mathbb{R}^3)^*)$, we can define the operator
\\
$A^*(\mu)\psi(s) = \begin{cases} -\frac{d\psi(s)}{ds} & s\in(0,1]\\
 \int_{-1}^{0} d\eta^{T}(t,0)\, \psi(-t)& s=0\end{cases}
$
\\
and the bilinear inner product
\begin{equation}\label{eq:bilinear}
\left \langle \psi(s),\phi(\theta) \right \rangle=\bar{\psi}(0)\phi(0)-\int_{-1}^{0} \int_{\xi=0}^{\theta} \bar{\psi}(\xi-\theta)d\eta(\theta)\phi(\xi)\, d\xi
\end{equation}
where $\eta(\theta)=\eta(\theta,0)$. Then $A(0)$ and $A^*$ are the adjoint operators with respect to the above bilinear form. By the discussion of the previous section, we know that $\pm i\nu_0\tau_c$ are eigenvalues of $A(0)$, thus they are also eigenvalues of $A^*$. We now need to compute the eigenvectors of $A(0)$ and $A^*$ corresponding to $+i\nu_0\tau_c$ and $-i\nu_0\tau_c$, respectively.
\\Suppose that $q(\theta)=(1,q_2,q_3)^{T}e^{i\theta \tau_c \nu_0}$, $\theta\in[-1,0]$ is the eigenvector of A(0) corresponding to $i\tau_c\nu_0$. Then, $A(0)q(\theta)=i\tau_c\nu_0 q(\theta)$. It follows from the definition of $A(0)$ and from $\ref{eq:L_mu},\ref{eq:L_integ},\ref{eq:eta}$ that
\begin{equation}
\tau_c
\begin{pmatrix}
i\nu_0+\sigma & -\sigma & 0 \\
(r+\frac{x^{2}_{r}}{b})(e^{-i\nu_0\tau_c}-1) & -\gamma+(\gamma+\sigma)e^{-i\nu_0\tau_c}+i\nu_0 & x_r(e^{-i\nu_0\tau_c}-1)\\
-x_r & -x_r & b+i\nu_0
\end{pmatrix}
\begin{pmatrix}
1\\
q_2\\
q_3
\end{pmatrix}
=\begin{pmatrix}
0\\
0\\
0
\end{pmatrix}
\end{equation}
We can easily obtain
\begin{equation}
q(\theta)=(1,q_2,q_3)^{T}=
\begin{pmatrix}
1,
\frac{i\nu_0+\sigma}{\sigma},
\frac{\sigma x_r+x_r(i\nu_0+\sigma)}{\sigma(b+i\nu_0)}
\end{pmatrix}^{T}e^{i\theta\nu_0\tau_c}
\end{equation}

Similarly, let $q^*(s)=D(1,q^*_2,q^*_3)^{T}e^{is \tau_c \nu_0}$, $s\in[0,1]$ , is the eigenvector of $A^{*}$ corresponding to  $-i\tau_c\nu_0$ and we find:
\begin{equation}
(1,q^*_2,q^*_3)
\begin{pmatrix}
i\nu_0-\sigma & \sigma & 0 \\
(r+\frac{x^{2}_{r}}{b})(1-e^{i\nu_0\tau_c}) & \gamma-(\gamma+\sigma)e^{i\nu_0\tau_c}+i\nu_0 & x_r(1-e^{i\nu_0\tau_c})\\
x_r & x_r & -b+i\nu_0
\end{pmatrix}
=(0,0,0)
\end{equation}
We can easily obtain
\begin{equation}
q^{*}(s)=D
\begin{pmatrix}
1,
\frac{b(i\nu_0-\sigma)(b-i\nu_0)}{(1-e^{i\nu_0\tau_c})[2bx_r^2+rb^2-i\nu_0(rb+x_r^2)]},
\frac{b(i\nu_0-\sigma)}{2bx_r^2+rb^2-i\nu_0(rb+x_r^2)}
\end{pmatrix}
e^{is\nu_0\tau_c}
\end{equation}

The 'orthonormality' condition $\left \langle q^*(s),q(\theta) \right \rangle=1$, helps us determining the value of $D$. By the definition $\ref{eq:bilinear}$ we have


\begin{equation}
D=\frac{1}{1+q^*_2\bar{q_2}+q^*_3\bar{q_3}-\tau_c e^{i\tau_c\nu_0}[(r+\frac{x_r^2}{b})q^*_2+(\gamma+\sigma)\bar{q_2}q^*_2+x_r q^*_2\bar{q_3}]}.
\end{equation}
Furthermore, we also have that $\left \langle q^*(s),\bar{q}(\theta) \right \rangle=0$.
\\
The next step is to compute the coordinates to describe the center manifold $\bf{C}_0$ at $\mu=0$. Let $x_t$ be a solution of Eq.$\ref{eq:vdot}$ when $\mu=0$. We define 
\begin{equation}\label{eq:Wz}
z(t)=\left \langle q^*,x_t \right\rangle, \qquad W(t,\theta)=x_t(\theta)-2Re\{z(t)q(\theta)\}
\end{equation}
On the center manifold $\bf{C}_0$ we have $W(t,\theta)=W(z(t),\bar{z}(t),\theta)$, where
\begin{equation}
W(z,\bar{z},\theta)=W_{20}(\theta)\frac{z^2}{2}+W_{11}(\theta)z\bar{z}+W_{02}(\theta)\frac{\bar{z}^2}{2}+W_{30}(\theta)\frac{z^3}{6}+\dots,
\end{equation}
and $z$ and $\bar{z}$ are the coordinates for the center manifold in the direction of $q$ and $\bar{q}$. Since $W$ is real if $x_t$ is real, we consider only real solutions. For $x_t\in\bf{C}_0$, since $\nu=0$, we have
\begin{equation}
\begin{split} 
\dot{z}(t) & =i\tau_c\nu_0z+\bar{q}^*(\theta)f(0,W(z,\bar{z},\theta)+2Re\{zq(\theta)\})\\
&=i\tau_c\nu_0z+\bar{q}^*(0)f(0,W(z,\bar{z},0)+2Re\{zq(0)\})=i\tau_c\nu_0z+\bar{q}^*(0)f_0(z,\bar{z})\\
&=i\tau_c\nu_0z+g(z,\bar{z})\end{split}
\end{equation}
with
\begin{equation}\label{eq:g_svil}
g(z,\bar{z})=g_{20}\frac{z^2}{2}+g_{11}z\bar{z}+g_{02}\frac{\bar{z}^2}{2}+g_{21}\frac{z^2\bar{z}}{2}+\dots
\end{equation}
Noticing $x_t(\theta)=(x_{1t}(\theta),x_{2t}(\theta),x_{3t}(\theta))=zq(\theta)+\bar{zq}(\theta)+W(t,\theta)$, we have
\vspace{0.1cm}\\
$
x_{1t}(0)=z+\bar{z}+W_{20}^{(1)}(0)\frac{z^2}{2}+W_{11}^{(1)}(0)z\bar{z}+W_{02}^{(1)}(0)\frac{\bar{z}^2}{2}+O(\left |(z,\bar{z})\right|^3),
$\vspace{0.1cm}\\
$
x_{2t}(0)=q_{2}z+\bar{q_{2}}\bar{z}+W_{20}^{(2)}(0)\frac{z^2}{2}+W_{11}^{(2)}(0)z\bar{z}+W_{02}^{(2)}(0)\frac{\bar{z}^2}{2}+O(\left |(z,\bar{z})\right|^3),
$\vspace{0.1cm}\\
$
x_{3t}(0)=q_{3}z+\bar{q_{3}}\bar{z}+W_{20}^{(3)}(0)\frac{z^2}{2}+W_{11}^{(3)}(0)z\bar{z}+W_{02}^{(3)}(0)\frac{\bar{z}^2}{2}+O(\left |(z,\bar{z})\right|^3),
$
\vspace{0.1cm}\\
and similarly we have for $x_{1t}(-1),x_{2t}(-1),x_{3t}(-1)$. 

\vspace{0.3cm}
Thus, from Eq.$\ref{eq:g_svil}$ we get
\begin{equation}
g(z,\bar{z})=\bar{q}^*(0)f_0(z,\bar{z})=\bar{D}(1,\bar{q_2}^*,\bar{q_3}^*)\tau_c
\begin{pmatrix}
0\\
-x_{1t}(0)x_{3t}(0)+x_{1t}(-1)x_{3t}(-1)\\
x_{1t}(0)x_{2t}(0)
\end{pmatrix}
\end{equation}


Comparing the coefficients with those of Eq.$\ref{eq:g_svil}$, we obtain
\\

\begin{flalign*}
&g_{20}=2\bar{D}\tau_c(\bar{q_2}^* q_3 e^{-2i\tau_c\nu_0}-\bar{q_2}^* q_3+\bar{q_3}^* q_2)\\
&g_{11}=2\bar{D}\tau_c\bar{q_3}^* Re(q_2)\\
&g_{02}=2\bar{D}\tau_c(\bar{q_2}^* \bar{q_3} e^{2i\tau_c\nu_0}-\bar{q_2}^* \bar{q_3}+\bar{q_3}^* \bar{q_2})
\end{flalign*}

\begin{equation}
g_{21}= -2\bar{q_2}^*\bar{D}\tau_c\left[\frac{1}{2}\bar{q_3}W_{20}^{(1)}(0)+q_3 W_{11}^{(1)}(0)+\frac{1}{2}W_{20}^{(3)}(0)+W_{11}^{(3)}(0)\right]+
\end{equation}

\begin{flalign*}
&+2\bar{q_2}^*\bar{D}\tau_c\left[W_{11}^{(3)}(-1)e^{-i\tau_c\nu_0}+\frac{1}{2}W_{20}^{(3)}(-1)e^{i\tau_c\nu_0}+W_{11}^{(1)}(-1)q_3 e^{-i\tau_c\nu_0}+\frac{1}{2}W_{20}^{(1)}(-1)\bar{q_3}e^{i\tau_c\nu_0}\right]+\\
&+2\bar{q_3}^*\bar{D}\tau_c\left[W_{11}^{(2)}(0)+\frac{1}{2}W_{20}^{(2)}(0)+W_{11}^{(1)}(0)q_2 +\frac{1}{2}W_{20}^{(1)}(0)\bar{q_2}\right].
\end{flalign*}

Since there are $W_{11}^{(j)}(\theta)$ and $W_{20}^{(j)}(\theta)$ in $g_{21}$, we still need to compute them. From $\ref{eq:vdot_second}$ and $\ref{eq:Wz}$ we have:
\begin{equation}\label{eq:W_dot}
\begin{split}
\dot{W}=&\dot{x_t}-\dot{z}q-\dot{\bar{z}}\bar{q}\\
=&\begin{cases} AW-2Re\{\bar{q}^*(0)f_oq(\theta)\} & \theta\in[-1,0)\\
 AW-2Re\{\bar{q}^*(0)f_oq(0)\}+f_0 & \theta=0\end{cases}\\
=&AW+H(z,\bar{z},\theta)
\end{split}
\end{equation}
where
\begin{equation}\label{eq:H_sviluppo}
H(z,\bar{z},\theta)=H_{20}(\theta)\frac{z^2}{2}+H_{11}(\theta)z\bar{z}+H_{02}(\theta)\frac{\bar{z}^2}{2}+\dots
\end{equation}
Expanding the above series and comparing the corrisponding coefficients, we get
\begin{equation}\label{eq:AWH}
(A-2i\tau_c\nu_0)W_{20}(\theta)=-H_{20}(\theta)\qquad AW_{11}(\theta)=-H_{11}(\theta),\dots
\end{equation}
From Eq. $\ref{eq:W_dot}$ we know that for $\theta\in[-1,0)$,
\begin{equation}
H(z,\bar{z},\theta)=-\bar{q}^*(0)f_0q(\theta)-q^*(0)\bar{f}_0\bar{q}(\theta)=-gq(\theta)-\bar{g}\bar{q}(\theta).
\end{equation}
Comparing the coefficients with those in Eq.$\ref{eq:H_sviluppo}$ gives that
\begin{equation}\label{eq:H20}
H_{20}(\theta)=-g_{20}q(\theta)-\bar{g}_{02}\bar{q}(\theta),
\end{equation}
and
\begin{equation}\label{eq:H11}
H_{11}(\theta)=-g_{11}q(\theta)-\bar{g}_{11}\bar{q}(\theta).
\end{equation}
From $\ref{eq:AWH}$ and $\ref{eq:H20}$ and the definition of A, it follows that 
\begin{equation}
\dot{W}_{20}(\theta)=2i\nu_0\tau_c W_{20}(\theta)+g_{20}q(\theta)+\bar{g}_{02}\bar{q}(\theta).
\end{equation}
Hence
\begin{equation}\label{eq:W20}
{W}_{20}(\theta)=\frac{ig_{20}}{\nu_0\tau_c}q(0)e^{i\theta\tau_c\nu_0}+\frac{i\bar{g}_{02}}{3\nu_0\tau_c}\bar{q}(0)e^{-i\theta\tau_c\nu_0}+E_1e^{2i\theta\tau_c\nu_0}
\end{equation}
and similarly, from $\ref{eq:AWH}$ and $\ref{eq:H11}$ we get
\begin{equation}\label{eq:W11}
{W}_{11}(\theta)=-\frac{ig_{11}}{\nu_0\tau_c}q(0)e^{i\theta\tau_c\nu_0}+\frac{i\bar{g}_{11}}{\nu_0\tau_c}\bar{q}(0)e^{-i\theta\tau_c\nu_0}+E_2
\end{equation}
where $E_1=(E_1^{1},E_1^{2},E_1^{3}),E_2=(E_2^{1},E_2^{2},E_2^{3})\in\mathbb{R}^3$ are constant vectors.
Now it remains to determine appropriate values for $E_1$ and $E_2$. From the definition of A and $\ref{eq:AWH}$, we obtain
\begin{equation}\label{eq:intW20}
\int_{-1}^{0} d\eta(\theta)\,W_{20}(\theta)=2i\nu_0\tau_c W_{20}(0)-H_{20}(0)
\end{equation}
and
\begin{equation}\label{eq:intW11}
\int_{-1}^{0} d\eta(\theta)\,W_{11}(\theta)=-H_{11}(0).
\end{equation}

\begin{figure*}[h]
\begin{center}
\includegraphics[width=12cm,height=5cm]{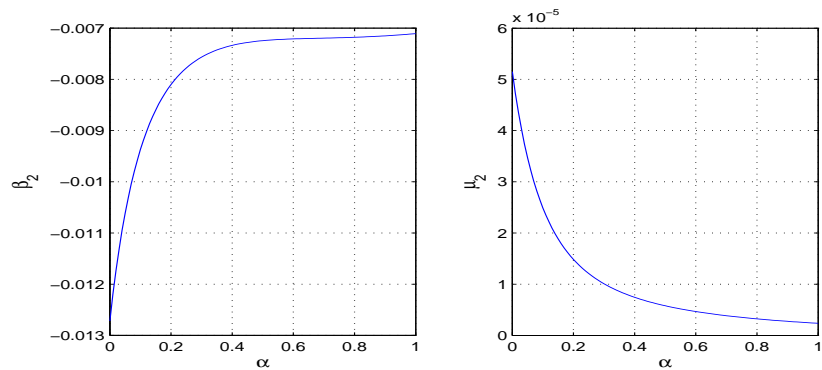}
\end{center}
\caption{Plots of (a) $\beta_2$ and (b) $\mu_2$ as functions of $\alpha$.}
\label{fig:beta_mu_alpha}
\end{figure*}

By $\ref{eq:W_dot}$ we have:
\begin{equation}\label{eq:H20solve}
H_{20}(0)=-g_{20}q(0)-\bar{q}_{02}\bar{q}(0)+2\tau_c\begin{pmatrix}
0\\
-q_3+q_3e^{-2i\tau_c\nu_0}\\
q_2\end{pmatrix}
\end{equation}
and
\begin{equation}\label{eq:H11solve}
H_{11}(0)=-g_{11}q(0)-\bar{q}_{11}\bar{q}(0)+2\tau_c\begin{pmatrix}
0\\
0\\
Re(q_2)
\end{pmatrix}
\end{equation}
Substituting $\ref{eq:W20}$ and $\ref{eq:H20}$ into $\ref{eq:intW20}$ and noticing that
\begin{center}
$
\left(i\tau_c\nu_0 I-\int_{-1}^{0} e^{i\theta\tau_c\nu_0}\, d\eta(\theta)\right)q(0)=0
$
\end{center}
and
\begin{center}
$
\left(-i\tau_c\nu_0 I-\int_{-1}^{0} e^{-i\theta\tau_c\nu_0}\, d\eta(\theta)\right)\bar{q(0)}=0
$
\end{center}
we obtain
\begin{center}
$
\left(2i\tau_c\nu_0 I-\int_{-1}^{0} e^{2i\theta\tau_c\nu_0}\, d\eta(\theta)\right)E_1=2\tau_c\begin{pmatrix}
0\\
-q_3+q_3e^{-2i\tau_c\nu_0}\\
q_2
\end{pmatrix}
$
\end{center}
and similarly, substituting $\ref{eq:W11}$ and $\ref{eq:H11}$ into $\ref{eq:intW11}$ we can get:
\begin{center}
$E_1=2G^{-1}\begin{pmatrix}
0\\
-q_3+q_3e^{-2i\tau_c\nu_0}\\
q_2
\end{pmatrix},
$\hspace{0.3cm}
$E_2=2G'^{-1}\begin{pmatrix}
0\\
0\\
Re(q_2)
\end{pmatrix}
$\end{center}
where
\\
$
G=\begin{pmatrix}
2i\nu_0+\sigma & -\sigma & 0\\
(e^{-2i\tau_c\nu_0}-1)(r+\frac{x_r^2}{b}) & 2i\nu_0-\gamma+(\gamma+\sigma)e^{-2i\tau_c\nu_0} & x_r(e^{-2i\tau_c\nu_0}-1)\\
-x_r & -x_r & b+2i\nu_0\end{pmatrix},
G'=\begin{pmatrix}
\sigma & -\sigma & 0\\
0 & \sigma & 0\\
-x_r & -x_r & b\end{pmatrix}
$ 
Thus, we can determine the coefficients $W_{20}(0)$,$W_{11}(0)$ and $g_{21}$. Therefore, each $g_{ij}$ id determined by the parameters and the delay in $\ref{eq:vdot}$. So, we can compute the following values:
\begin{equation}
c_1(0)=\frac{i}{2\nu_0}\left[g_{20}g_{11}-2\left|g_{11}\right|^2-\frac{1}{3}\left|g_{02}\right|^2\right]+\frac{g_{21}}{2},
\qquad
\mu_2=-\frac{Re\{c_1(0)\}}{Re\{\lambda '(\tau_c)\}}
\end{equation}
\begin{equation}
\beta_2=2Re\{c_1(0)\}\qquad T_2=-\frac{Im\{c_1(0)\}+\mu_2Im\{\lambda '(\tau_c)\}}{\tau_c\nu_0}.
\end{equation}
which determine the quantities of bifurcating periodic solutions in the center manifold at the critical value of delay $\tau_c$. The sign of $\mu_2$ determines the direction of Hopf bifurcation: if $\mu_2>0$ $(\mu_2<0)$, then the bifurcating periodic solutions exist for $\tau>\tau_c$ $(\tau<\tau_c)$ and the bifurcation is supercritical (subcritical). The quantity $\beta_2$ determines the stability of the bifurcating periodic solutions, i.e. they are stable (unstable) for $\beta_2<0$ $(\beta_2>0)$. 
\begin{theorem}
Assume that conditions of Theorem $\ref{transv_cond}$ hold. Then
\begin{enumerate} 
\item if $Re\{c_1(0)\}<0$, then there exist periodic solutions bifurcating from $E_{+}$ for $\tau>\tau_c$, and they are orbitally asymptotically stable as $t\shortrightarrow\infty$;
\item if $Re\{c_1(0)\}>0$, then there exist periodic solutions bifurcating from $E_{+}$ for $\tau<\tau_c$, and they are orbitally asymptotically stable as $t\shortrightarrow-\infty$.
\end{enumerate}
\end{theorem} 
\textit{Remark}. Though the complexity of the expression of $Re\{c_1(0)\}$ will not allow a direct study of its sign as a function of $\alpha$, we show the numerical simulations performed: $\beta_2$ is negative for all $\alpha\in\left[0,1\right]$ (see Fig.$\ref{fig:beta_mu_alpha}$ (a)), whereas $\mu_2$ is a positive function of $\alpha$ on the interval $\left[0,1\right]$ (see Fig.$\ref{fig:beta_mu_alpha}$ (b)), as aspected because $Re\{\lambda '(\tau_c)\}>0$. So we can conclude that the bifurcation arising from system $\ref{eq:gener_control}$ is supercritical and the periodic solutions are stable for all $\alpha\in\left[0,1\right]$ .

\section{Conclusions}
In this paper we have considered the whole family of the generalized Lorenz system in its chaotic regime. This system bridges the gap between the Lorenz and the Chen systems, and by varying the parameter $\alpha$ on the interval $\left[0,1\right]$ we obtain the particular system of the family. We have used a feedback technique with a non linear control to achieve the stabilization of the steady states and we have carried out a mathematical analysis of the global dynamics of the system and studied the dependence on the time delay and the characteristic parameter $\alpha$, also investigating the existence of stability switches. We have also shown that the time delay can destabilize the steady states and lead to periodic solutions through Hopf bifurcation. By using the  the normal form theory and center manifold argument, we determine direction, stability and period of these periodic solutions and also prove that they are stable and the bifurcation is supercritical for all $\alpha\in\left[0,1\right]$.

\end{document}